\documentclass{aa} 
\usepackage{comment}
\usepackage[labelfont=bf]{caption}
\usepackage[colorlinks = true,
            linkcolor = blue,
            urlcolor  = blue,
            citecolor = blue,
            anchorcolor = blue]{hyperref}
\usepackage{multirow}
\usepackage{natbib}
\usepackage{soul}
\bibpunct{(}{)}{;}{a}{}{,} 
\usepackage{graphicx}
\usepackage[toc]{appendix}
\usepackage{verbatim}
\usepackage{txfonts}
\begin{document}

   \title{Detection of H$\alpha$ emission from PZ Tel B using SPHERE/ZIMPOL\thanks{ESO programme ID: 0101.C-0672(A); P.I.: A. Musso Barcucci. The reduced data products are available via ESO/phase 3.}}

   \author{Arianna Musso Barcucci
          \inst{1}
          \and Gabriele Cugno\inst{2}
          \and Ralf Launhardt\inst{1}
          \and Andr\'e M\"uller\inst{1}
          \and Judit Szulagyi\inst{3}
          \and Roy van Boekel\inst{1}
          \and Thomas Henning\inst{1}
          \and Mickael Bonnefoy\inst{4}
          \and Sascha P. Quanz\inst{2}
          \and Faustine Cantalloube\inst{1}
          }

   \institute{Max Planck Institute for Astronomy (MPIA),
              K\"onigstuhl 17, 69117 Heidelberg, Germany\\
              \email{musso@mpia.de}
              \and
        ETH Z\"urich, Institute for Particle Physics and Astrophysics, Wolfgang-Pauli-Str. 27, 8093 Z\"urich, Switzerland
        \and
        Center for Theoretical Astrophysics and Cosmology, Institute for Computational Science, University of Z\"urich, Winterthurerstrasse 190, CH-8057 Z\"urich, Switzerland
        \and
        Univ. Grenoble Alpes, CNRS, IPAG, 38000 Grenoble, France}

   \date{Received --; accepted --}
\definecolor{green2}{rgb}{0,0.8,0.2}
\abstract{
H$\alpha$ is a powerful tracer of accretion and chromospheric activity, which has been detected in the case of young brown dwarfs and even recently in planetary mass companions (e.g. PDS70\,b and c).
H$\alpha$ detections and characterisation of brown dwarf and planet companions can further our knowledge of their formation and evolution, and expanding such a sample is therefore our primary goal. We used the Zurich Imaging POLarimeter (ZIMPOL) of the SPHERE instrument at the Very Large Telescope (VLT) to observe the known $38-72$M$_{\mathrm{J}}$ companion orbiting PZ Tel, obtaining simultaneous angular differential imaging observations in both continuum and narrow H$\alpha$ band. We detect H$\alpha$ emission from the companion, making this only the second H$\alpha$ detection of a companion using the SPHERE instrument. We used our newly added astrometric measurements to update the orbital analysis of PZ Tel B, and we used our photometric measurements to evaluate the H$\alpha$ line 
flux. Given the estimated bolometric luminosity, we obtained an H$\alpha$ activity (log$\mathrm{(L_{H\alpha}/L_{bol}})$) between $-4.16$ and $-4.31$. The H$\alpha$ activity of PZ Tel B is consistent with known average activity levels for M dwarf of the same spectral type. Given the absence of a known gaseous disk and the relatively old age of the system (24 Myr), we conclude that the H$\alpha$ emission around PZ Tel B is likely due to chromospheric activity.} 


   \maketitle
%

\section{Introduction}
H$\alpha$ emission from low-mass stars and brown dwarfs can have multiple origins. In the case of young objects ($<10$ Myr) gas from the circumstellar disk can be accreted onto a circumsecondary disk and, due to the high temperatures of the shock front, this can lead to dissociation of H2 molecules and consequent H$\alpha$ emission \mbox{\citep{Szulagyi_2017,Aoyama_2018}}.
In the case of young non-accreting stars, chromospheric activity produces well-known emission lines, with H$\alpha$ being one of the most prominent ones.

H$\alpha$ emission from single low-mass stars and brown dwarfs has been extensively studied through the years. \mbox{\cite{West_2004}} used around 8000 single M dwarf spectra from the Sloan Digital Sky Survey (SDSS) to evaluate their H$\alpha$ flux and investigate the activity fraction and strength as a function of spectral type.
They quantified the activity as logarithm of the ratio between the H$\alpha$ luminosity and bolometric luminosity, and they found a peak in the fraction of active stars around spectral type M8, where more than 70$\%$ of stars were active. They also evaluated the mean activity strength as the ratio between the H$\alpha$ luminosity, and the bolometric luminosity finding that it is constant between M0 and M5 and that it declines at later spectral types.
Similar trends were recovered by subsequent surveys: \mbox{\cite{Lee_2009}} studied the short-term H$\alpha$ variability of 43 single M dwarf, finding a similar decrease in the activity strength for later spectral types, as well as an increase in the variability level up to spectral type M7. \mbox{\cite{Kruse_2010}} also focused on short-timescale H$\alpha$ variability using nearly $53000$ spectra from SDSS; they recovered both the log$(L_{H\alpha}/L_{\mathrm{bol}})$ activity trend (that increases until $\sim$M6 with subsequent decrease) and the same variability trend (that increases with later spectral type). More recently, \mbox{\cite{Robertson_2013}} studied the correlation between activity, mass, spectral type, and metallicity of 93 stars ranging from K5 to M5. They find that the activity trend is recovered and, at a given stellar mass, metal rich stars appear to be more active.

However, much less is known about the H$\alpha$ emission from companions in binary systems, the main reason being the difficulty in disentangling the two components in the spectrum (with few exceptions, see e.g. \mbox{\citealt{Bowler_2014}}, \mbox{\citealt{Santamaria_2018}}).
Few remarkable H$\alpha$ detections, often associated with accretion, have been made using high-contrast imaging techniques, which allow to differentiate between the two components in a binary system and evaluate the H$\alpha$ flux from the companion. One example is HD\,142527\,B, an accreting M-dwarf companion first detected in H$\alpha$ by \mbox{\cite{Close_2014}} with the Magellan Adaptive Optics system (MagAO).
The companion is later re-detected using the Zurich Imaging POLarimeter (ZIMPOL) of the SPHERE instrument at the Very Large Telescope (VLT) by \mbox{\cite{Cugno_2019}}, who also searched for local accretion signals in other objects suspected of hosting forming giant planets. More recently, \mbox{\cite{Wagner_2018}} claim the detection of H$\alpha$ emission from the young planet PDS\,70\,b. \mbox{\cite{Haffert_2019}} were also able to detect H$\alpha$ emission from PDS\,70\,b with the MUSE Integral Field Spectrograph at the VLT \mbox{\citep{Bacon_2010}} and identified another accreting protoplanet in the same system, PDS\,70\,c.
\mbox{\cite{Sallum_2015}} claimed to have detected accretion from the companion orbiting around LkCa\,15, but recent studies from \mbox{\cite{Thalmann_2016}} and \mbox{\cite{Currie_2019}} could not confirm it, also doubting whether the companions exist at all. Other remarkable H$\alpha$ detections include GQ Lup b and DH Tau b, both detected by \mbox{\cite{Zhou_2014}} using the Hubble Space Telescope, and three newly detected brown dwarf companions from the Upper Sco region \citep{Petrus_2019}.

These detections are fundamental for various reasons: firstly, they prove that it is feasible to detect planets and low-mass stellar companions using H$\alpha$ emission as a tracer; secondly, they give initial insight into the gas-accretion phase of planet and brown dwarf formation; and thirdly, they show that it is possible to use state of the art high contrast imaging instruments and techniques to detect H$\alpha$ emission in binary systems.\\
In order to learn more about the early stages of planet formation and evolution, increasing the number of directly imaged known companions with H$\alpha$ detection is our primary goal.
In this work, we present SPHERE/ZIMPOL angular differential imaging (ADI, \citealt{adi_marois2006}) observations in H$\alpha$ of the known companion orbiting around the star PZ Tel. In Section 2 we present the target and in Section 3 we detail the observations and data reduction; we present the analysis and the results in Section 4 and we summarise our conclusions in Section 5.
\begin{table}[t!]
\caption{Fundamental parameters and properties of the PZ\,Tel system.}             
\label{table: basic} 
\centering  
\begin{tabular}{lcc}
\hline \hline
Parameter       & \multicolumn{2}{c}{value}    \\ \hline
RA [hh:mm:ss]   & \multicolumn{2}{c}{+18:53:05.87} \\
DEC [dd:mm:ss]  & \multicolumn{2}{c}{-50:10:49.90} \\
Parallax [mas]  & \multicolumn{2}{c}{$21.2186 \pm 0.0602$ \tablefootmark{a}}\\
Distance [pc]   & \multicolumn{2}{c}{$47.13 \pm 0.13$}  \\
Age [Myr]   & \multicolumn{2}{c}{$24\pm3$\tablefootmark{b}}                               \\
A$_V$ [mag] & \multicolumn{2}{c}{$0.53^{+0.84}_{-0.53}\,\tablefootmark{c}$}\\ \hline
                        & PZ Tel A      & PZ Tel B     \\
Sp. Type   & G6.5V\tablefootmark{c}  & M7$\pm$1  \\
T$_\mathrm{eff}$ [K]    & $\sim5338 \pm 200$\tablefootmark{b} & 2500-2700\tablefootmark{c}\tablefootmark{d}\tablefootmark{e} \\
Mass & $1.13 \pm 0.03\,\mathrm{M}_\odot$\tablefootmark{b} & 38-72 M$_\mathrm{J}\tablefootmark{e}$\\
${[}\mathrm{Fe}/\mathrm{H}{]}$ ${[}\mathrm{dex}{]}$ & $0.05 \pm 0.20$\tablefootmark{b}  & $0.30_{-0.30}\tablefootmark{c}$\\ 
$\mathrm{v\,sin\,i}$ [km/s] & $73 \pm 5$\tablefootmark{b}  & --- \\ 
\multirow{2}{*}{L ${[}\mathrm{L}_{\odot}{]}$}& \multirow{2}{*}{$1.16 \pm 0.1\tablefootmark{e}$} & $0.002^{+0.0004}_{-0.0003}\tablefootmark{c}$\\
    & & $0.003\pm0.0008\tablefootmark{e}$ \\
\hline
\end{tabular}
\tablebib{\tablefoottext{a}{\cite{Gaia_2018}.}\tablefoottext{b}{\cite{Jenkins_2012}.}\tablefoottext{c}{\cite{Schmidt_2014}; due to the model grid used, it is not possible to place an upper limit on the companion's metallicity.}\tablefoottext{d}{\cite{Mugrauer_2010}.}\tablefoottext{e}{\cite{Maire_2016}}}
\end{table}
\section{PZ\,Tel B}

\begin{table}[t!]
\caption{Summary of observations and detector characteristics.}             
\label{table: observations} 
\centering  
\begin{tabular}{lcc}
\hline \hline
Parameter & \multicolumn{2}{c}{Value} \\
\hline
\multicolumn{3}{c}{Observational setup}\\
\hline
Observation date & \multicolumn{2}{c}{30/05/2018}\\
Run ID & \multicolumn{2}{c}{0101.C-0672(A)} \\
$\#$ Science frames & \multicolumn{2}{c}{20}\\
$\#$ Flux frames & \multicolumn{2}{c}{8}\\ 
$\#$ Centre frames & \multicolumn{2}{c}{4}\\ 
DIT Science ${[}\mathrm{s}{]}^{a}$ & \multicolumn{2}{c}{220} \\
DIT Flux ${[}\mathrm{s}{]}^{b}$ & \multicolumn{2}{c}{52} \\
Tot. time ${[}\mathrm{min}{]}^{c}$ & \multicolumn{2}{c}{73.3} \\
Flux time ${[}\mathrm{min}{]}^{d}$ & \multicolumn{2}{c}{6.9} \\
Seeing ${[}\mathrm{arcsec}{]}^{e}$ & \multicolumn{2}{c}{0.9} \\
Tot. field rotation ${[}\mathrm{deg}{]}^{f}$& \multicolumn{2}{c}{50.05} \\
Platescale ${[}\mathrm{mas/pix}{]}$ & \multicolumn{2}{c}{3.6$\times$3.6} \\
Coronagraph & \multicolumn{2}{c}{V\_CLC\_M\_WF} \\
\hline
\multicolumn{3}{c}{ZIMPOL detector characteristics}\\
\hline
                                    & Cnt\_H$\alpha$ & N\_H$\alpha$ \\
$\lambda_{0}\,{[}\mathrm{nm}{]}^{g}$ & 644.9 & 656.34 \\
$\Delta \lambda {[}\mathrm{nm}{]}^{h}$ & 3.83 & 0.75 \\
Cnt. Zp. $\mathrm{{[}erg/cm^{2}/ADU/A{]}^{i}}$ & $1.59^{+0.05}_{-0.05}\times10^{-17}$ &  $10^{+0.05}_{-0.05}\times10^{-17}$\\
Line Zp. $\mathrm{{[}erg/cm^{2}/ADU{]}^{j}}$ & --- & $9.2^{+4}_{-0.5}\times10^{-16}$\\
\hline
\end{tabular}
\caption*{$^{a}$Detector Integration Time for the science observations; $^{b}$Detector Integration Time for the flux observations; $^{c}$Total on source integration time for the science frames; $^{d}$Total on source integration time for the flux frames; $^{e}$Median seeing throughout the observations; $^{f}$Total field rotation; $^{g}$Filter central wavelength; $^{h}$Filter equivalent width; $^{i}$Continuum zeropoints from \mbox{\cite{Schmid_2017}}; $^{j}$Line zeropoints from \mbox{\cite{Schmid_2017}}.}
\end{table}

PZ Tel (HD\,174429, HIP\,92680) is a G6.5 type star with an age of $24\pm3$ Myr (\mbox{\citealt{Jenkins_2012}}, \mbox{\citealt{Bell_2015}}), belonging to the $\beta$ Pic moving group \mbox{\citep{Zuckerman_2001}} at a distance of $\sim$47 pc \mbox{\citep{Gaia_2018}}. In 2010, two independent studies discovered a sub-stellar companion at a separation of $\sim$0.3 arcsec: \mbox{\cite{Mugrauer_2010}} with the NaCo instrument at the VLT and \mbox{\cite{Biller_2010}} with the Near-Infrared Coronagraphic Imager (NICI) at Gemini South. Both authors interpolated low-mass objects evolutionary tracks \mbox{\citep{Baraffe_2002}} and inferred a mass of \mbox{$28^{+12}_{-4}\,\mathrm{M_{J}}$} and \mbox{$36\pm6\,\mathrm{M_{J}}$}, which corresponds to a spectral type of M5-9. Following its discovery, the PZ Tel system has been the subject of several studies. \mbox{\cite{Jenkins_2012}} use spectra obtained with the Fiber-fed Extended Range Optical Spectrograph (FEROS)
to derive a rotational velocity of the host star of $\mathrm{v\,sin\,i=73\pm5\,km\,s^{-1}}$, a metallicity of $\mathrm{[Fe/H]=0.05\pm0.20\,dex}$, and an age of 5-27 Myr which led to a revised mass for PZ Tel B of $62\pm9\,\mathrm{M_{J}}$ via comparison with evolutionary models. Additional spectroscopic information was obtained with the Spectrograph for INtegral Field Observations in the Near Infrared (SINFONI) at the VLT (\mbox{\citealt{Schmidt_2014}}), leading to a mass estimate for the companion of $\mathrm{M}=7.5^{+16.9}_{-4.3}\,\mathrm{M_{J}}$, and a bolometric luminosity of ${\mathrm{log(L_{bol}/L_{\odot}}})=-2.66^{+0.06}_{-0.08}$, which are independent of both age and evolutionary model used. More recently, \mbox{\cite{Maire_2016}} obtained multi-band photometric observations of the companion using the InfraRed Dual-band Imager and Spectrograph (IRDIS), the Integral Field Spectrograph (IFS), and ZIMPOL at VLT/SPHERE, and derived a mass of 38-72 M$_\mathrm{J}$ (spectral type M7$\pm$1), which we use in this work. 
The observed mean activity strength value for this spectral type are -4.31 according to \mbox{\cite{West_2004}}, and -4.37 according to \mbox{\cite{Kruse_2010}}.
\mbox{\cite{Maire_2016}} also derived a bolometric luminosity for the companion of \mbox{${\mathrm{log(L_{bol}/L_{\odot}}})=-2.51\pm0.10$}.
\begin{figure*}[t!]
	\centering
		\includegraphics[width=\textwidth]{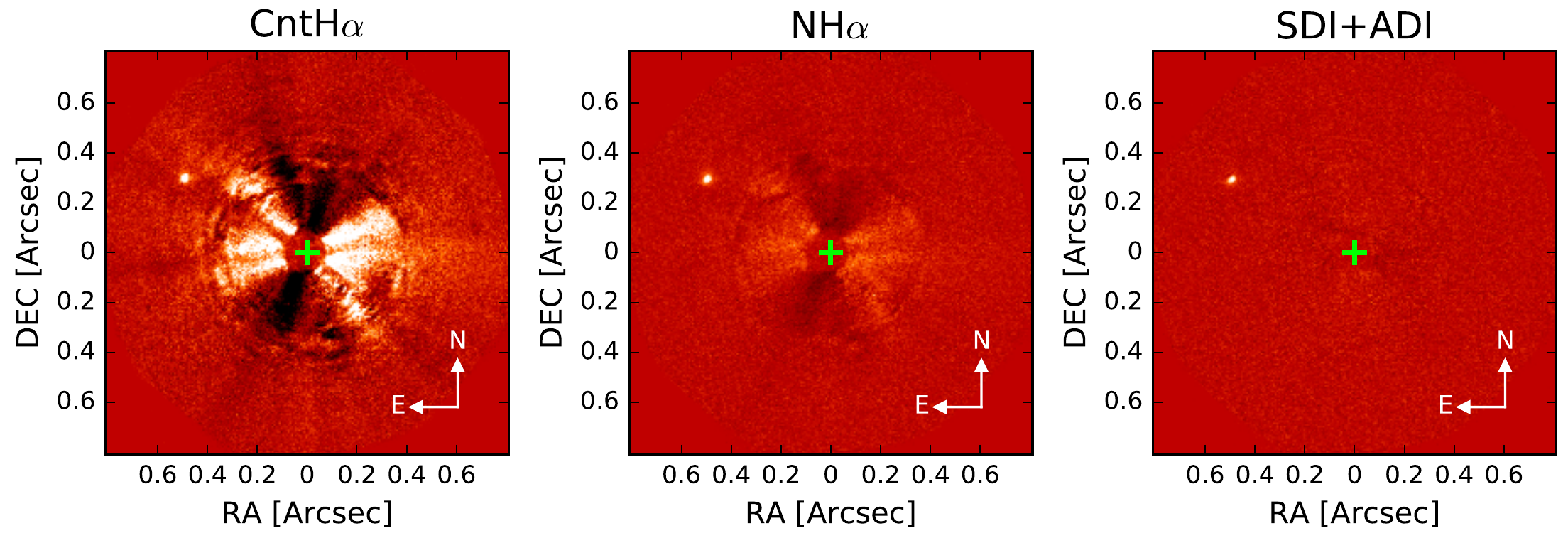}
		\caption{Reduced images showing PZ\,Tel\,B. In all three images the green cross marks the position of the central star, the data is oriented with north up, and the companion is clearly visible NE of the star. The images are normalised and the colour map was chosen for a better visualisation of the data. \textit{Left panel}: classical ADI reduced image of the continuum H$\alpha$ filter frames. \textit{Central panel}: classical ADI reduced image of the narrow band H$\alpha$ filter frames.
		\textit{Right panel}: ASDI analysis.}
	\label{Fig: detection}\
\end{figure*}
\mbox{\cite{Riviere-Marichalar_2014}} obtained Herschel-PACS far-IR photometric observations at 70, 100, and 160 $\mathrm{\mu m}$ of 19 $\beta$ Pic moving group members. They were able to exclude the presence of a substantial debris disk around PZ\,Tel, due to the non-detection of excess in the aforementioned bands, placing an upper limit on the infrared excess of \mbox{$\mathrm{L_{IR}/L_{\star}}<2.3\times10^{-5}$}.
\mbox{\autoref{table: basic}} summarises the host star and companion properties.

\section{Observations and data reduction}
We observed the PZ Tel system with the ZIMPOL instrument at VLT/SPHERE (\mbox{\citealt{Schmid_2018}}), obtaining simultaneous coronagraphic ADI observations in the Cnt\_H$\alpha$ and N\_H$\alpha$ filter. 
The data were taken on UT 2018-05-30 in two observation blocks before and after the meridian passage, to maximise the total field rotation while allowing flexibility in the observing schedule. Each observing block consists of a set of science exposures with an integration time of 220 seconds, which were bracketed with non-saturated observations of the star with DIT=52 seconds, that we denote as flux frames. We also recorded a centre frame at the beginning and end of each observing block, in which a pattern is applied to the deformable mirror creating 4 bright copies of the central PSF outside of the coronagraph (in a symmetric pattern around the central star) which are used to compute the stellar position behind the coronagraph.
The conditions were clear throughout the entire observations, with a median DIMM seeing of $0.9$ arcsec. Standard bias, dark and flat calibrations were observed on the same night. Details of the observations, as well as main ZIMPOL detector characteristics, are summarised in \mbox{\autoref{table: observations}}.\\
\indent The data was reduced using the ZIMPOL reduction pipeline developed and maintained at ETH Z\"urich which consists of: flat fielding, bias correction and dark subtraction, remapping the initial 7.2 $\times$ 3.6 mas/pix platescale into the squared grid of 3.6 $\times$ 3.6 mas/pix, and separating the frames in the two filters.
The pipeline was applied to the flux, centre, and science frames. To account for possible shifting of the stellar position on the detector during the observations, we fitted a two dimensional gaussian to each spot in the 4 cosmetically reduced centring frames (and in both filters), computing the centre as intersection of the connecting lines. The final centre and relative error are the mean and standard deviation of these 4 centres (for each filter). We then re-centre the science frames using the scipy.ndimage.interpolation.shift package with spline interpolation of order 3, and cut them to stamps of 1.62$\times$1.62 arcsec ending up with 20 cosmetically reduced and centred science frames for each filter.
Since the unsaturated star is offset from the coronagraph and therefore visible, we fitted a two dimensional gaussian to re-centre the flux frames, ending up with eight cosmetically reduced and centred flux frames for each filter.
The parallactic angle for each frame is automatically computed by the ZIMPOL pipeline, and takes care of a constant known offset of $134\pm0.5^{\circ}$ \mbox{\citep{Maire_2016, Cugno_2019}} for which the frames must be rotated in the counterclockwise direction.
\section{Analysis and results}
The goal of this paper is to detect and quantify H$\alpha$ emission from the companion around PZ Tel, to expand the sample of known brown dwarfs and planetary companions with H$\alpha$ detection and better understand the formation and evolution of these objects.
We also provide an additional astrometric measurement of the PZ Tel B, extending the time baseline by four more years. We clearly detected the companion in both Cnt\_H$\alpha$ and N\_H$\alpha$ filter, as shown in \mbox{\autoref{Fig: detection}}. Even though the detection is clear in both filters, we also analysed the data with angular spectral differential imaging (ASDI) technique, shown in the rightmost panel of \mbox{\autoref{Fig: detection}}. We refer to Appendix A, as well as \mbox{\cite{Cugno_2019}}, for a detailed explanation of the ASDI analysis.
\subsection{Astrometry and flux contrast}
We quantified the astrometry and flux contrast of the companion, for both filters, using the \textit{ANDROMEDA} package \mbox{\citep{Cantalloube_2015}}. This algorithm needs as input the cosmetically reduced frames, the corresponding parallactic angles, and an unsaturated PSF of the central star to create a model of the planetary signal signature, whose flux and position is fitted via a maximum likelihood estimation. We created this unsaturated image of the host star (for both filters) as median combination of all the flux frames, scaled to the DIT of the science frames. We set the Inner Working Angle parameter to $\mathrm{1.0\,\lambda/D}$ (we refer to the \textit{ANDROMEDA} paper for a detailed explanation of how the package works).
The astrometry and flux contrast evaluated with \textit{ANDROMEDA} are presented in \mbox{\autoref{tab:andromeda results}}.
\begin{table}[]
    \centering
    \caption{Astrometry and flux contrast evaluated with \textit{ANDROMEDA}, for both continuum and narrow band filter.}
    \begin{tabular}{c| c c}
    \hline \hline
         Parameter & Cnt\_H$\alpha$ & N\_H$\alpha$\\
         \hline
         Sep. ${[}\mathrm{arcsec}{]}$ &  0.5666$\pm$0.0036  & 0.5669$\pm$0.0036\\
         P.A. ${[}\mathrm{deg}{]}$ & 58.93$\pm$0.5 &  59.40$\pm$0.5 \\
         Flux contrast & $(7.4\pm0.9)\times10^{-5}$ & $(29.0\pm 3.5)\times10^{-5}$ \\
         \hline
    \end{tabular}
    \label{tab:andromeda results}
\end{table}
\begin{table*}[t!]
    \centering
    \caption{Astrometric measurements for PZ Tel B available in the literature}
    \begin{tabular}{c c c c c c}
    \hline
    \hline
    Epoch & Separation & P.A. & Instrument & Filter & ref. \\
    \hline
         2007/06/13 & $255.6\pm2.5$ & $61.68\pm0.6$ & NaCo & Ks & \cite{Mugrauer_2012}\\
        2009/04/11 & $330.0\pm10$& $59.0\pm1.0$ & NICI & CH$_4$ 4\% Long+Short & \cite{Biller_2010}\\
         2009/09/28 & $336.6\pm1.2$ & $60.52\pm0.22$ & NaCo & Ks & \cite{Mugrauer_2012}\\
         2010/05/05-07$^{a}$ & $355.7\pm0.8$ & $60.33\pm0.14$ & NaCo & J/H/Ks & \cite{Mugrauer_2012}\\
         2010/05/09 & $360.0\pm3.0$ & $59.4\pm0.5$ & NICI & CH$_4$ 1\% Short & \cite{Biller_2010}\\
         2010/09/26 & $365.0\pm8.0$ & $59.2\pm0.8$ & NaCo & $L'$ & \cite{Beust_2016}\\
         2010/10/28  & $369.3\pm1.1$ & $59.91\pm0.18$ & NaCo & Ks & \cite{Mugrauer_2012}\\
         2011/03/25  & $382.2\pm1.0$ & $59.84\pm0.19$ & NaCo & Ks & \cite{Mugrauer_2012}\\
         2011/04/24 & $373.0\pm9.0$ & $58.7\pm0.2$ & NICI & CH$_4$ 4\% Long+Short & \cite{Biller_2013}\\
         2011/05/03 & $394.0\pm2.0$ & $60.4\pm0.2$ & NaCo & Ks & \cite{Beust_2016}\\
         2011/06/03-06$^{a}$  & $388.3\pm0.5$ & $59.69\pm0.10$ & NaCo & Ks & \cite{Mugrauer_2012}\\
         2011/06/07 & $390.0\pm5.0$ & $60.0\pm0.6$ & NaCo & $L'$ & \cite{Beust_2016}\\
         2012/04/05 & $397.0\pm9.0$ & $60.4\pm0.2$ & NICI & CH$_4$ 4\% Long+Short & \cite{Biller_2013}\\
         2012/06/08$^{a}$  & $419.4\pm0.6$ & $59.58\pm0.03$ & NaCo & Ks & \cite{Ginski_2014}\\
         2014/07/13$^{a}$   & $477.46\pm1.02$ & $59.82\pm0.24$ & SPHERE/IRDIS & H2$\&$H3 & \cite{Maire_2016}\\
         2014/08/07$^{a}$   & $479.65\pm0.034$ & $59.94\pm0.23$ & SPHERE/IRDIS &  K1$\&$K2 & \cite{Maire_2016}\\
         \hline
    \end{tabular}
    \caption*{ $^{(a)}$The quoted values are weighted means of several values presented in the cited papers.}
    \label{tab:astrometry from literature}
\end{table*}
\subsection{Photometry}
We followed the prescriptions from \mbox{\cite{Cugno_2019}} to convert the flux contrasts into physical fluxes for both filters. The only difference was that, due to the presence of the coronagraph, we evaluated the flux of the host star using the flux frames instead of the science frames. 
Given the vicinity of the bands, we assumed that the continuum flux density is the same in both filters and we evaluated the flux in both. We then used the continuum flux density to evaluate the contamination of the narrow band filter due to continuum emission, and we corrected for it, obtaining the H$\alpha$ line flux.
We refer to Appendix B for a detailed step-by-step description of the analysis (we also performed an alternative photometric analysis described in Appendix C).
After correcting for extinction (see Appendix B and \mbox{\autoref{table: basic}}), the total flux in the continuum filter, the total flux in the narrow band filter and the line flux, are:\\
$F_{\mathrm{Cnt\_Ha}}^{\star}=(5.68\pm0.18)\times10^{-11}\,\mathrm{erg/cm^{2}/s}$\\
$F_{\mathrm{N\_Ha}}^{\star}=(1.47\pm0.09)\times10^{-11}\,\mathrm{erg/cm^{2}/s}$\\
$F_{\mathrm{N\_Ha,line}}^{\star}=(3.53\pm0.8)\times10^{-12}\,\mathrm{erg/cm^{2}/s}$.\\
\begin{figure}[t!]
	\centering
		\includegraphics[width=\hsize]{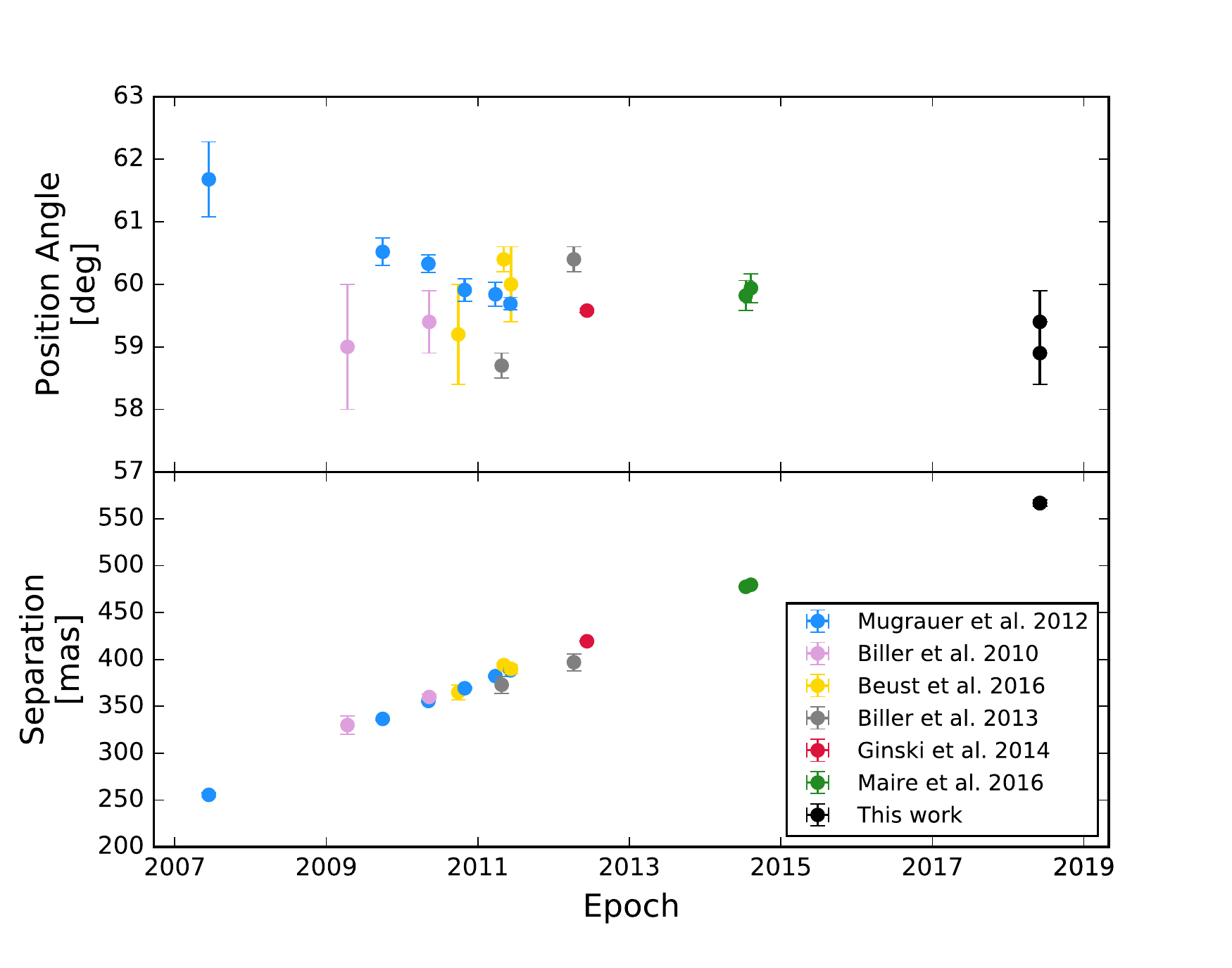}
		\caption{Separation and position angles of PZ Tel B, at various epochs. We show the astrometric values found in the literature (see \mbox{\autoref{tab:astrometry from literature}}) in various colours, together with the astrometry presented in this paper, for both filters (black points).}
	\label{Fig: astrometry}
\end{figure}
\begin{figure}[t!]
	\centering
		\includegraphics[width=\hsize]{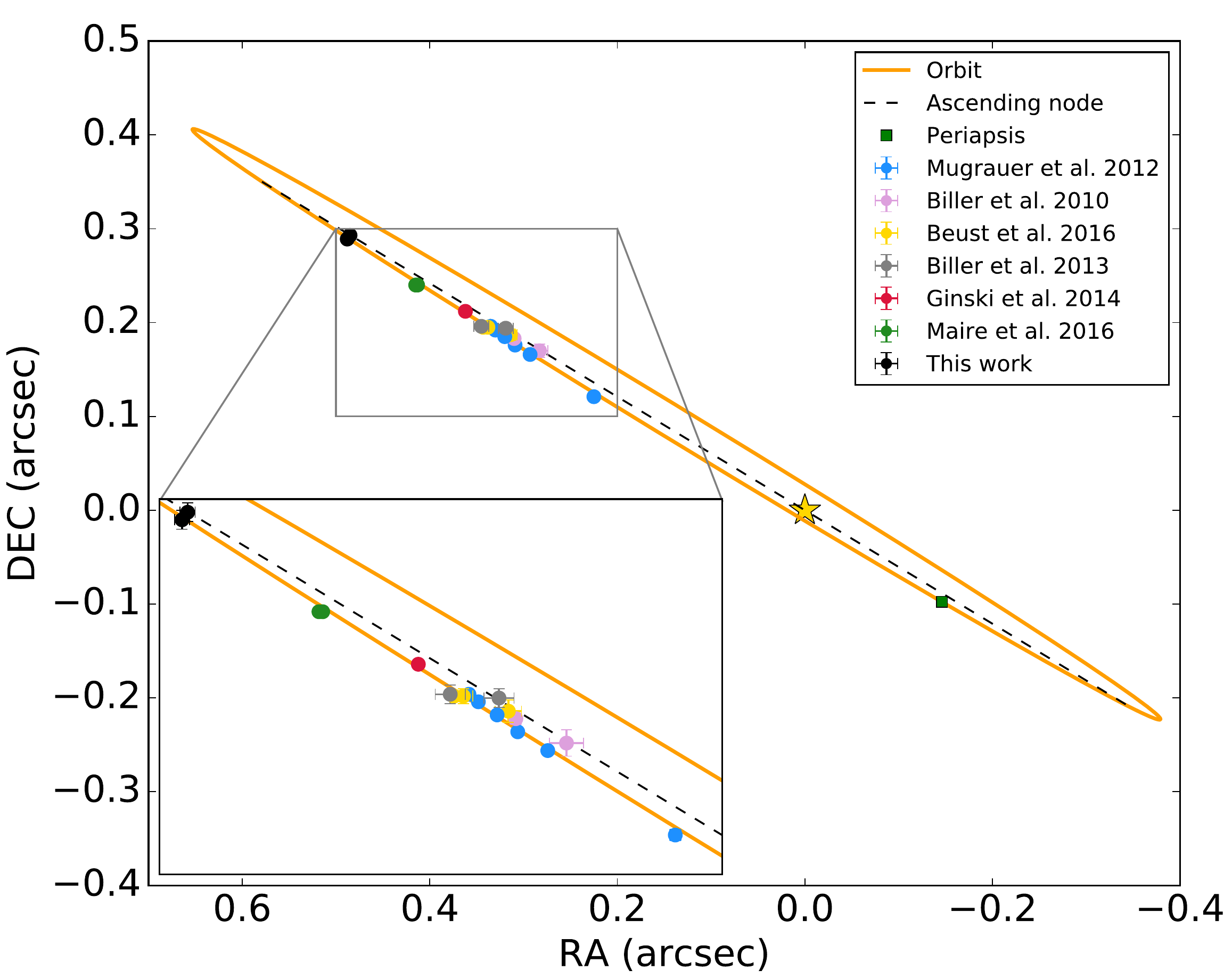}
		\caption{Best orbital solution found with PyAstrOFit. The orange line shows the orbit, the yellow star marks the position of the primary and the various astrometric measurements are shown with the same colour coding of \mbox{\autoref{Fig: astrometry}}. The green square marks the position of the periapsis, and the dashed black line shows the position of the ascending node.}
	\label{Fig: bestorbit}
\end{figure}

We now have the flux of the primary in the two filters and, together with the companion flux contrast (see \mbox{\autoref{tab:andromeda results}}), we can calculate the companion flux in both bands. The companion line flux is then the difference between the fluxes in the two filters (normalising the continuum flux to the width of the H$\alpha$ filter). The final values for the companion are:
$F_{\mathrm{Cnt\_Ha}}^B=(1.92\pm0.9)\times10^{-15}\,\mathrm{erg/cm^{2}/s}$\\
$F_{\mathrm{N\_Ha}}^B=(2.54\pm0.8)\times10^{-15}\,\mathrm{erg/cm^{2}/s}$\\
$F_{\mathrm{Ha\_line}}^B=(2.17\pm0.9)\times10^{-15}\,\mathrm{erg/cm^{2}/s}$.\\

The companion H$\alpha$ line flux can be converted into a luminosity, multiplying by the squared distance, obtaining:\\
\mbox{$\mathrm{L}_{H\alpha}=(1.51\pm0.05)\times10^{-7}\,\mathrm{L_{\odot}}$}.

Finally, we can evaluate the H$\alpha$ activity as the ratio between the H$\alpha$ luminosity and the bolometric luminosity of the object. For a bolometric luminosity of PZ Tel B of \mbox{${\mathrm{log_{10}(L_{bol}}}/\mathrm{L_{\odot}})$=$-2.66^{+0.06}_{-0.08}$} (\mbox{\citealt{Schmidt_2014}}), we obtain an H$\alpha$ activity of \mbox{$\mathrm{log_{10}(L_{H\alpha}/L_{bol}})$=$-4.16\pm0.08$}. 
Similarly, we obtain \mbox{$\mathrm{log_{10}(L_{H\alpha}/L_{bol}})$=$-4.31\pm0.1$} in the case of \mbox{$\mathrm{log_{10}(L_{bol}/L_{\odot}})$=$-2.51\pm0.10$} (\mbox{\citealt{Maire_2016}}). The H$\alpha$ activity values agree within the errorbars.
\subsection{Orbital constraints}
Following its discovery in 2010, PZ Tel B has been observed several times in the last years, providing various astrometric measurements on an increasingly large time baseline. We compiled all the available astrometric datapoints from the literature in \mbox{\autoref{tab:astrometry from literature}} and we show the position angle and separation of the companion through time in \mbox{\autoref{Fig: astrometry}}. With our newly added observations, the available baseline is now $\sim$12 years.
\mbox{\cite{Mugrauer_2012}} were the first to report a deceleration of the variation of the the angular separation of the companion, to be expected for an object moving on a Keplerian orbit towards apastron, which would support a bound orbit solution. Deceleration was also detected by \mbox{\cite{Ginski_2014}} and \mbox{\cite{Maire_2016}}. We revisited the literature data and, together with our newly added astrometry, we further confirm this trend. The angular separation increases with a rate of \mbox{$\mathrm{d_{sep}/t}=35.3\pm1.2\,\mathrm{mas/yr}$} between June 2007 and September 2009, and then of $32.9\pm1.6\,\mathrm{mas/yr}$ between September 2009 and September 2010.
The rate keeps decreasing all the way down to $27.7\pm0.6\,\mathrm{mas/yr}$ between June 2012 and July 2014 and, finally, of just $23.0\pm0.3\,\mathrm{mas/yr}$ between July 2014 and May 2018. Given the deceleration of the companion, we decided to restrict the following orbital analysis to bound orbits only ($e\leq1$).

Given the newly extended astrometric baseline, we explored the possible orbital solutions using the Python package PyAstrOFit\footnote{\url{https://github.com/vortex-exoplanet/PyAstrOFit}} \mbox{\citep{Wertz_2017}} which provides a series of tools to fit orbits using the emcee package \mbox{\citep{Foreman-Mackey_2013}} with the modified Markov chain Monte Carlo (MCMC) approach described in \mbox{\cite{Goodman_Weare_2010}}.
We assumed uniform prior distribution for the semi-major axis ($a$), the eccentricity ($e$), the inclination ($i$), the longitude of ascending node ($\Omega$) and argument of periapsis ($\omega$), and the time of periastron passage ($t_\mathrm{p}$).
Assuming a system mass of $1.2\,\mathrm{M_{\odot}}$ (see \mbox{\autoref{table: basic}}), we explored all possible bound solutions ($e\leq1$), allowing a range of semi-major axis between 10 and 1200 au, an inclination between 10 and 180 degrees, and $\Omega$ and $\omega$ within natural boundaries. The only other hyperparameters are the number of walkers (which we set to 1200), and the scale parameter $a$, which directly impacts the acceptance rate AR \mbox{\citep{Mackay_2003}} of the walkers. We manually tuned $a$ to ensure an AR between 0.2 and 0.5. PyAstrOFit relies on the Gelman Rubin $\hat{R}$ statistical test to check for convergence \mbox{\citep{Gelman_1992,Ford_2006}}, which is considered reached when all the parameters pass the test three times in a row (with a threshold of $\hat{R}\textless 1.05$, where the closer the $\hat{R}$ value is to 1 the closer the Markov chain is to convergence).

\begin{figure*}[t!]
	\centering
		\includegraphics[width=\textwidth]{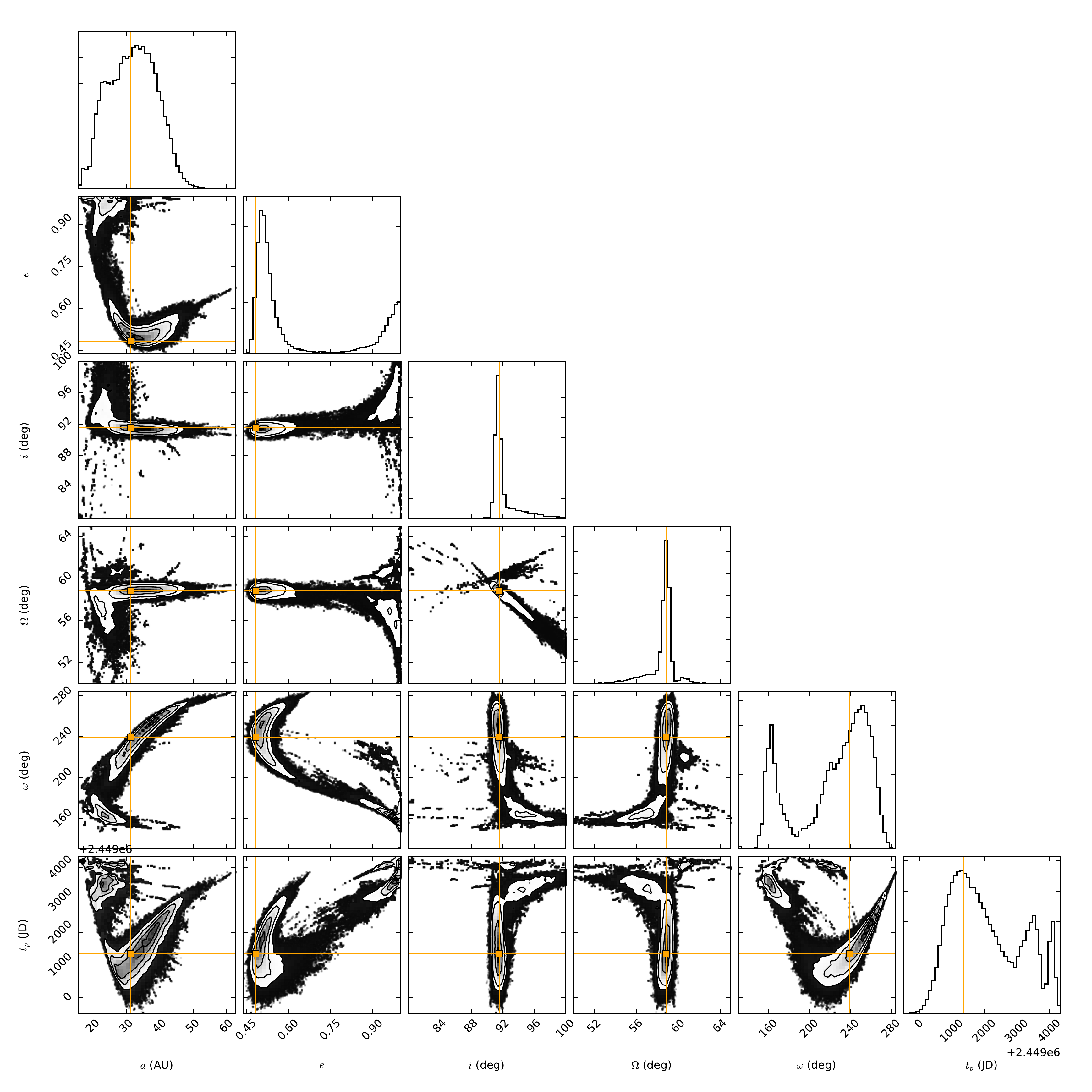}
		\caption{Posterior distributions of the orbital elements (\textit{diagonal panels}) and correlation between the parameters (\textit{off-axis panels}). The orange lines and squares mark the position of the best solution found in terms of reduced $\chi^{2}$, as reported in \mbox{\autoref{tab:MCMC results}}.}
	\label{Fig: cornerplot}\
\end{figure*}

The posterior distributions of the orbital elements, as well as the correlation between them, is shown in the corner plot of \mbox{\autoref{Fig: cornerplot}}. The eccentricity distribution shows two peaks at $\sim$0.55 and at 1, which is a lower boundary smaller than what found by previous studies (\mbox{$0.62<e<0.99$} in \mbox{\cite{Ginski_2014}} and $e\gtrsim0.66$ in \mbox{\cite{Maire_2016}}) and significantly smaller than the lower boundary of 0.91 found in the most recent orbital study of PZ Tel B, by \mbox{\cite{Beust_2016}}. 
A possible explanation for this difference lies in the different boundaries applied: \mbox{\cite{Beust_2016}} allowed not-bound orbits while in this work we only considered orbits with \mbox{$e\textless1$}.
Our best solution for the semi-major axis of 31.3 au agrees with previous works (\mbox{$17.86<a<1098$ au} in \mbox{\citealt{Ginski_2014}} and \mbox{$a\gtrsim24.5$ au} in \mbox{\citealt{Maire_2016}}). We found a best inclination of 91.6 degrees, which is in agreement with previous ranges of \mbox{$91.3^{\circ}<i<168.1^{\circ}$} for \mbox{\cite{Ginski_2014}} and \mbox{$91^{\circ}<i<96.1^{\circ}$} for \mbox{\cite{Maire_2016}}. Previous confidence intervals for the longitude of ascending node were \mbox{$50^{\circ}<\Omega<70^{\circ}$} for \mbox{\cite{Ginski_2014}} and \mbox{$55.1^{\circ}<\Omega<59.1^{\circ}$} for \mbox{\cite{Maire_2016}}, and \mbox{\cite{Ginski_2014}} cited an interval of \mbox{$122.2^{\circ}<\omega<306^{\circ}$} for the argument of periapsis. All of these agree with our best solutions of \mbox{$\Omega=58.8^{\circ}$} and \mbox{$\omega=239.2^{\circ}$}. The best solution for the time of periastron passage corresponds to 1996.3, which agrees within the confidence intervals of previous works, but it is systematically lower than their best solutions (2002.9 for \mbox{\citealt{Mugrauer_2012}}, 2003.5 for \mbox{\citealt{Ginski_2014}} and 2002.5 for \mbox{\citealt{Beust_2016}}).

The best solutions in terms of reduced $\chi^{2}$ and the $1-\sigma$ confidence intervals are reported in \mbox{\autoref{tab:MCMC results}}. The orbit corresponding to these best parameters is shown in \mbox{\autoref{Fig: bestorbit}}, where we overplot the astrometric points (both from literature and from this work) as well as the position of the host star, the direction of the ascending node and the position of periapsis.\\
Our new astrometric datapoints are in agreement with previous measurements in terms of orbital elements, and help to tighten the uncertainties.
\begin{table}[t!]
    \centering
    \caption{Best solutions in terms of reduced $\chi^{2}$ and $1-\sigma$ confidence intervals for all the orbital elements. These orbital elements have an associated $\chi^{2}_{\mathrm{red}}$ of 2.15.}
    \begin{tabular}{l| c}
    \hline \hline
         Parameter & Value\\
         \hline
         a [AU] & [21.4,39.9]\\
         a$_{\chi^{2}}$ & 31.3\\
         e & [0.48,0.99]\\
         e$_{\chi^{2}}$ & 0.48\\
         i [deg] & [90.7,92.1]\\
         i$_{\chi^{2}}$ & 91.6\\
         $\Omega$ [deg] & [58.3,59.3]\\
         $\Omega_{\chi^{2}}$ & 58.8\\
         $\omega$ [deg] & [155.2,265.6]\\
         $\omega_{\chi^{2}}$ & 239.2\\
         t$_{p}$ [MJD] & [49655.2, 53206.3]\\
         $t_{p,\chi^{2}}$ & 50346.6\\
         \hline
    \end{tabular}
    \label{tab:MCMC results}
\end{table} 
\section{Discussion and conclusions}
We presented SPHERE/ZIMPOL observations of the known sub-stellar M dwarf companion around PZ Tel, taken in both H$\alpha$ continuum and narrow band filter. We detected the companion in both datasets obtaining new astrometric and photometric measurements. This currently represents the second only H$\alpha$ detection of a companion using the SPHERE instrument, and it further proves the capability of this instrument to detect H$\alpha$ signatures in binary systems.\\
\indent We used our newly added astrometric data, together with values from the literature, to explore the allowed orbital solutions for PZ Tel B, finding orbital elements in agreement with what done in previous works (with the only exception being our lower boundary on the eccentricity). Our added data extends the available baseline for orbital studies of PZ Tel B up to $\sim$12 years. We find that the companion is clearly decelerating over time, which is to be expected for a Keplerian bounded object moving towards apastron.
We evaluated the H$\alpha$ luminosity and activity of PZ Tel B, finding values for  $\mathrm{log_{10}(L_{H\alpha}/L_{bol}})$ of -4.16$\pm$0.08 and -4.31$\pm$0.10, for bolometric luminosities of -2.66 and -2.51, respectively.\\
\indent Several studies investigated the H$\alpha$ activity in M dwarf, both as a function of spectral type and mass. \mbox{\cite{West_2004}} evaluated the average H$\alpha$ activity as a function of spectral type, finding an average activity of -4.0, -4.31 and -4.10 for spectral types of M6, M7 and M8, respectively. A later study from \mbox{\cite{Kruse_2010}} found similar average activity levels of -3.89, -4.35 and -4.17 for the same spectral types.
Based on its spectral type, PZ Tel B thus appears to be slightly less active than the average, while still be consistent with the average values within the uncertainties.\\
\indent Given the age of the system, and the absence of a known gaseous disk, it is unlikely that the observed H$\alpha$ luminosity is due to accretion processes. The fact that the activity level is consistent with what is expected for an object of spectral type M6-8, leads us to conclude that the most likely explanation for the H$\alpha$ luminosity observed in PZ Tel B is chromospheric activity.

Finally, we suggest that a possible explanation for the below average H$\alpha$ value of PZ Tel B is that the object has a variable emission and we happened to observe it during a moment of low activity. This reasoning is supported primarily by the late spectral type of the object, which is known to correlate with a higher variability level (see, e.g. \mbox{\citealt{Kruse_2010}}); in addition, the companion has a high metallicity, which \mbox{\cite{Robertson_2013}} correlated with a higher activity. However, follow-up H$\alpha$ observations would be needed to establish whether PZ Tel B displays a variable chromospheric activity.

\begin{acknowledgements}
Based on observations collected at the European Southern Observatory under ESO programme 0101.C-0672(A).
A.M.B. would like to thank the anonymous referee for the useful and constructive comments.
J.\,S. acknowledges the support from the Swiss National Science Foundation (SNSF) Ambizione grant PZ00P2\_174115.
A.M. acknowledges the support of the DFG priority programme SPP 1992 'Exploring the Diversity of Extrasolar Planets' (MU 4172/1-1 and QU 113/6-1). 
G.C. and S.P.Q. thank the Swiss National Science Foundation for the financial support under the grant number 200021$\_$169131.
This research made use of Astropy,\footnote{http://www.astropy.org} a community-developed core Python package for Astronomy \mbox{\citep{Astropy_2013, Astropy_2018}}.
\end{acknowledgements}


\bibliographystyle{aa}
\bibliography{Bibliography.bib}
\clearpage
\begin{appendices}
\newpage
\section*{Appendix A - Angular Spectral Differential Imaging}
The ASDI technique is a two step combination of spectral differentual imaging (SDI) \citep{Racine_1999_sdi} and ADI tecnique, where the images are first reduced with the SDI method, and then combined with a classical ADI reduction. The SDI tecnique relies on comparing images taken in different wavelengths, since any physical object would maintain the same position while speckles and Airy patterns would scale and move radially as a function of wavelength. In order to compare the continuum frames to the narrow band filter frames, we modified the continuum images as follows: we multiply all the Cnt\_Ha frames by the ratio of the NHa filter width to the Cnt\_Ha filter width (see \mbox{\autoref{table: observations}}), in order to correct for the different filter throughput.\\ We then stretch these normalised Cnt\_Ha frames radially, by the ratio of the filters central wavelengths, using spline interpolation. This step is done in order to align the speckle patterns.\\
We subtracted these modified Cnt\_Ha frames to the NHa frames, in order to correct for all the wavelength-dipendent patterns.\\
We finally reduced these subtracted frames using classical ADI reduction (the frames are de-rotated to the same parallactic angle and median combined) producing the ASDI reduced image shown in the right panel of \mbox{\autoref{Fig: detection}}.

\section*{Appendix B - Photometry}
We follow the prescription in \cite{Cugno_2019} , Section 4.1.4, but applying it to the flux frames, because the science frames have a coronagraph blocking the central star.
\\For the extinction calculation, we use the extinction law of \cite{Cardelli_1989}:\\
$A_{\lambda}=a(\lambda)+\frac{b(\lambda)}{R_{V}}$\\
With $a(\lambda)$ and $b(\lambda)$ interpolated at $\lambda\sim0.65\,\mu m$ ($a(\lambda)=0.91$ and $b(\lambda)=-0.26$), $R_{V}=3.1$ and $A_{V}=0.53^{+0.84}_{-0.53}$ from \mbox{\cite{Schmidt_2014}}; obtaining $A_{H\alpha}=0.44^{+0.69}_{-0.44}$. We use the value of 0.44, without uncertainties.\\
We proceeded as follows: in the flux frames part of the pixels are obscured due to the spider and the coronagraph. We manually create a mask over these features and interpolate the flux frames using the interpolate.griddata package of $scipy$, with a linear interpolation.\\
We calculate the count rate in the single flux frames inside an aperture of radius 1.3 arcsec, using the $photutils$ Python package to create the desired aperture and sum all the pixel values inside (the package allows for fraction of pixels to be taken int o account).
Due to the relative low integration time for the flux frames (52 seconds) the frames are read-out noise dominated, rather than background dominated. To account for this, we also evaluated the count rates in a background annulus around the central star and, scaling according to the area, we subtracted the background counts to the total counts.
We do this for both continuum and narrow band frames.
We then evaluate the mean count rate and relative uncertainty $\sigma /\sqrt{n}$ and divide them by the integration time, obtaining the count rate per second $cts_{CntHa}$=$70353.6\pm258.1$, and $cts_{NHa}$=$14094.0\pm80.4$.\\
We convert these count rates into flux densities using eq. 1 of \cite{Cugno_2019} or eq.4 of \cite{Schmid_2017}, as:
\begin{equation}
F_{\lambda}^{\star}=cts \times 10^{0.4(am\,k_{1} + m_{mode})} \times c^{cont}_{zp}
\end{equation}
With $am$ being the airmass during the observations, $k_{1}$ being the atmospheric extinction correction at Paranal ($0.085\pm0.004$ for Cnt\_Ha and $0.081\pm0.002$ for N\_Ha, from \citealt{Patat_2011}), and $c^{cont}_{zp}$ being the zeropoint for the desired filter (see \mbox{\autoref{table: observations}}).\\
So, the flux density in the Continuum filter Cnt\_Ha, is: $F_{\lambda}^{\star}(\mathrm{Cnt\_Ha})=(9.9\pm0.3)\times10^{-13}\,\mathrm{erg/cm^{2}/s/A}$.\\

We assume that the flux density of the primary is the same in both continuum and narrow band filter. We then calculate the flux in the continuum filter $F_{\mathrm{Cnt\_Ha}}$, and the flux in the narrow filter due to the continuum emission $F_{\mathrm{N\_Ha,cont}}$, as the continuum flux density multiplied by the two filter widths.
After correcting for the extinction, the two fluxes (in the continuum filter, and in the narrow filter due to the continuum emission) are:\\
$F_{\mathrm{Cnt\_Ha}}^{\star}=(5.68\pm0.18)\times10^{-11}\,\mathrm{erg/cm^{2}/s}$\\
$F_{\mathrm{N\_Ha,cont}}^{\star}=(1.11\pm0.04)\times10^{-11}\,\mathrm{erg/cm^{2}/s}$\\

The continuum flux density can also be used to estimate the counts in the narrow band filter that are due to the emission in the continuum, using eq.2 of \cite{Cugno_2019}. We obtain $cts_{NHa}=11186.2\pm665.9$ counts.\\
Subtracting these counts to the total counts evaluated in the N\_Ha filter (i.e: $cts_{NHa}$) allows us to obtain the counts in the filter due to line emission only, which are then converted into a line flux using eq.1 (with line zeropoint).
After correcting for extinction, we obtain:\\ $F_{\mathrm{N\_Ha,line}}^{\star}=(3.53\pm0.8)\times10^{-12}\,\mathrm{erg/cm^{2}/s}$.\\

The final total flux in the narrow filter is then the sum of the line and continuum contribution:\\
$F_{\mathrm{N\_Ha}}^{\star}=F_{\mathrm{N\_Ha,line}}^{\star}+F_{\mathrm{N\_Ha,cont}}^{\star}$\\$F_{\mathrm{N\_Ha}}^{\star}=(1.47\pm0.09)\times10^{-11}\,\mathrm{erg/cm^{2}/s}$.

\section*{Appendix C - Alternative Photometric Analysis}
We also performed the photometric analysis with an alternative method, which addresses the assumption that the flux density of the primary is the same in both filters.
We selected a suitable PHOENIX model spectrum \mbox{\citep{Husser_2013}} with the stellar parameters reported in \mbox{\autoref{table: basic}}.
We reduced publicly available FEROS spectrum of the primary, and used the aforementioned PHOENIX model to flux-calibrate them in units of $\mathrm{erg/s/cm^{2}/A}$.\\
We integrated the calibrated FEROS spectrum over the ZIMPOL filters, obtaining a synthetic photometry; which we then corrected comparing it the observed ZIMPOL photometry (see Appendix B). The resulting correction factors are 0.93 for the N\_Ha and 1.28 for the Cnt\_Ha filters, respectively.\\
We calculated the Cnt\_Ha to N\_Ha flux ratio. Now, instead of assuming that the flux density of the primary is the same in both filters, we use this filter flux ratio to correctly evaluate the continuum flux density of the primary in the N\_Ha filter.\\
We then use a PHOENIX model spectrum with the parameters of the PZ Tel B (see \mbox{\autoref{table: basic}}) to estimate its theoretical value in band fluxes. As expected, the Cnt\_Ha flux matches the observed one, while the measured N\_Ha flux is much brighter than the one expected from the model, due to the presence of H$\alpha$ emission.\\
We used the PHOENIX model of PZ Tel B to evaluate the flux ratio between the two filters, and then we used it to predict the continuum contribution to the measured N\_Ha flux based on the measured Cnt\_Ha flux.
Subtracting the continuum contribution to the N\_Ha flux leaves only the line contribution and, after accounting for the filter transmission curve, we obtain a H$\alpha$ line flux of $2.90\times10^{-15}\,\mathrm{erg/cm^{2}/s}$.\\

The H$\alpha$ line flux obtained with this alternative method is consistent within uncertainties with the value of \mbox{$(2.17\pm0.9)\times10^{-15}\,\mathrm{erg/cm^{2}/s}$} reported in Section 4.2.
We also evaluated the impact that a different PHOENIX model spectrum for PZ Tel B can have on the final results, assuming the lower and upper end of the parameters reported in \mbox{\autoref{table: basic}}. For a temperature of 2500 K, a bolometric luminosity of $0.002\,\mathrm{L_{\odot}}$ and a mass of 38 M$_\mathrm{J}$ we obtain a H$\alpha$ line flux of $2.90\times10^{-15}\,\mathrm{erg/cm^{2}/s}$. While for T$=$2700 K, L $=0.003\,\mathrm{L_{\odot}}$ and M $=$ 72 M$_\mathrm{J}$ we obtain a line flux of $2.90\times10^{-15}\,\mathrm{erg/cm^{2}/s}$. Both values agree with the the line flux reported in Section 4.2.

\end{appendices}
\end{document}